**Title:**

# The Evolving Ethics of Medical Data Stewardship

**Authors:**

Adam Leon Kesner, PhD, DABR
    Deputy Service Chief, Molecular and X-ray Imaging Physics
    Attending Physicist
    Memorial Sloan Kettering Cancer Center
    New York, NY, USA

Anyi Li, PhD
    Chief, Computer Service
    Memorial Sloan Kettering Cancer Center
    New York, NY, USA

Phillip Koo, MD, DABR
    Chief Medical Officer
    Prostate Cancer Foundation
    Los Angeles, CA, 10044

**Corresponding author:**
Adam L. Kesner, PhD, DABR
Attending Physicist, Memorial Sloan Kettering Cancer Center
New York, NY, USA
Email: kesnera@mskcc.org
Phone: +1-212-639-6371

# Abstract

Healthcare stands at a critical crossroads. Artificial Intelligence and modern computing are unlocking opportunities, yet their value lies in the data that fuels them. The value of healthcare data is no longer limited to individual patients. However, data stewardship and governance has not kept pace, and privacy-centric policies are hindering both innovation and patient protections. As healthcare moves toward a data-driven future, we must define reformed data stewardship that prioritizes patients' interests by proactively managing modern risks and opportunities while addressing key challenges in cost, efficacy, and accessibility.

Current healthcare data policies are rooted in 20th-century legislation shaped by outdated understandings of data—prioritizing perceived privacy over innovation and inclusion. While other industries thrive in a data-driven era, the evolution of medicine remains constrained by regulations that impose social rather than scientific boundaries. Large-scale aggregation is happening, but within opaque, closed systems. As we continue to uphold foundational ethical principles—autonomy, beneficence, nonmaleficence, and justice—there is a growing imperative to acknowledge they exist in evolving technological, social, and cultural realities.

Ethical principles should facilitate, rather than obstruct, dialogue on adapting to meet opportunities and address constraints in medical practice and healthcare delivery. The new ethics of data stewardship places patients first by defining governance that adapts to changing landscapes. It also rejects the legacy of treating perceived privacy as an unquestionable, guiding principle. By proactively redefining data stewardship norms, we can drive an era of medicine that promotes innovation, protects patients, and advances equity — ensuring future generations advance medical discovery and care.

# Introduction: All Moments Are Not Created Equal

"In effect, what we are experiencing now is the real-time development of property rights over personal data. It is the beginning of what will be a long regulatory journey, and the decisions being made now are setting the pathways for future economic, social, and political configurations."[1]

Healthcare data has not appreciably changed, but its value has. Following decades of Moore's law, its value proposition has evolved *exponentially*. This transformation is driven by advancements in computing infrastructure, artificial intelligence (AI), machine learning (ML), data analytics, and communications, allowing unprecedented processing power and scalability. These technologies support and thrive on large, diverse datasets, uncovering patterns and insights that were traditionally invisible to human researchers or traditional statistical methods. However, the true power of this innovation lies not in the computing itself but in the data that fuels it.[2,3]

The purpose of medicine has always been to link interventions with outcomes to improve patient care. Today, this foundational goal can be achieved in transformative ways through modern data processing infrastructures. Data fluidity has never been more consequential. A paradigm shift toward free-flowing healthcare data would enable a new era of innovation—one that moves beyond silos, connecting disparate sources of information to uncover patterns, enhance diagnostics, personalize treatments, reduce the cost of innovation, and inform public health strategies.

This shift would usher in a world where researchers and innovators no longer rely solely on limited, expensive, tightly controlled studies, but instead harness real-world data.[4] More accessible data would foster a healthcare ecosystem characterized by transparency, accessibility, and personalized risk management—elements the U.S. healthcare system currently lacks.[2] It would also open new, uncharted territory for protecting patients in ways that legacy systems and siloed data simply cannot.

Efforts to expand access to healthcare data quickly run up against a powerful obstacle: the deeply entrenched belief that privacy must take absolute precedence. Privacy is a value that deserves respect—but it should not be considered beyond reproach. Solutions and collective values evolve—and they are often multidimensional.[5] Privacy is one of several critical values we prioritize for patients—alongside innovation, treatment efficacy, accessibility, equity, and transparency. Moreover, privacy is less a fundamental principle than a mechanism for protecting against harm, and a world with evolving dangers requires evolving protections. And even if the notion of privacy is upheld as paramount, we must confront the reality that it no longer exists as we once understood it. In the US we operate with privacy centric policies in place, but with little checks, balances, or assurance. Most individuals are unaware of what medical data exists about them, where it is stored, or how it is used—entities like hospitals are collecting data, and data breaches are increasingly common, with little recourse available to affected individuals.[6] The intention behind privacy is laudable, but in contemporary practice, the 20th century legal infrastructure designed to enforce it undermines 21st century efforts for public-benefit, data-driven innovation, driving data collection and use into the shadows.

Consider the analogy of GPS navigation: by using these apps, individuals—often unknowingly—share their location data in a mutual value exchange. In return, industry provides free, competitive tools that offer real-time traffic updates, route optimization, and predictive travel times. While this involves a degree of privacy loss, the tradeoff is widely accepted because the benefits are immediate, tangible, and broadly distributed. This model not only enhances convenience but also reduces traffic congestion, emissions, and stress-related health impacts—offering meaningful benefits for both society and the environment. Yet similar innovation in healthcare remains constrained—not for lack of technological capacity, but due to data governance that treats privacy as sacrosanct, and does not even consider the expense of lost, potentially profound, collective benefit. This very same model could be applied to

healthcare—calibrating unintrusive real-time feedback based on time, place, and context to guide individuals toward beneficial decisions and away from harmful ones. Such guidance need not be limited to geography, but could extend to choices around diet, exercise, medication, sleep, mental health, environmental exposure, preventive care, and other behaviors that shape well-being over time—offering far more timely, personalized, and actionable advice than most patients currently receive, and at significantly lower cost to both individuals and society.

A similar contrast appears when we consider how we manage our personal memories versus our personal health. Cloud-based photo storage allows people to instantly access, search, and share a lifetime of images across devices—benefiting from intelligent algorithms that organize, tag, and curate their digital histories. Meanwhile, our medical records remain fragmented, inaccessible, and siloed—often even to our own physicians—leading to redundant testing, missed diagnostic connections, high expense and delays in care. This gap isn't just about convenience; it's about capability. The same kinds of personal photos and text that are quietly aggregated to train large language models now fuel generative tools reshaping industries. Medical data, if responsibly aggregated and shared, could power similarly transformative systems—expansive AI tools trained not just to chat, but to detect, predict, recommend, and guide. If driven by intentional openness, accessibility, and transparency, these tools could become significantly more valuable—more equitable, and more ethical—than the opaque systems currently being built by trillion-dollar industry players whose influence has no real counterbalance in shaping norms, markets, and even legislation.

In the US HIPAA driven healthcare data governance,[7] data privacy is supreme; protection models that once secured siloed data were appropriate for their time and served as anchors for 20th century medical innovation. Yet, it is these very policies that, when enacted in contemporary landscapes, continue to serve as anchors to outdated innovation models, hindering progress, escalating costs, and obstructing reduced cost/improved innovation opportunities - while simultaneously failing to provide the promised protection. Currently, healthcare spending represents more than 17% of U.S. GDP [8] and is a significant financial burden that has a major impact across society. Reports estimate that effectively leveraging healthcare data could save hundreds of billions of dollars while enhancing patient outcomes[9]—and that estimate excludes the added value of future innovations that could enable more personalized, lower-cost care at scale.

The power of exponential growth is often difficult to grasp when looking ahead. For example, a person taking 30 steps will travel roughly 30 meters—but if the distance doubles with each step, they will span the globe multiple times over. This simple illustration underscores a growing mismatch between the exponential rise of healthcare data and the linear pace of regulation and governance (see Figure 1). While regulatory-aligned data aggregation strategies tend to scale incrementally, healthcare data—and our ability to harness it—is expanding at an exponential rate.[10,11] Historically, the only systems that have consistently scaled innovation at exponential rates are those driven by open competition, incentives, and free-market dynamics. Infrastructure designed for linear growth will increasingly fall short of enabling meaningful innovation.

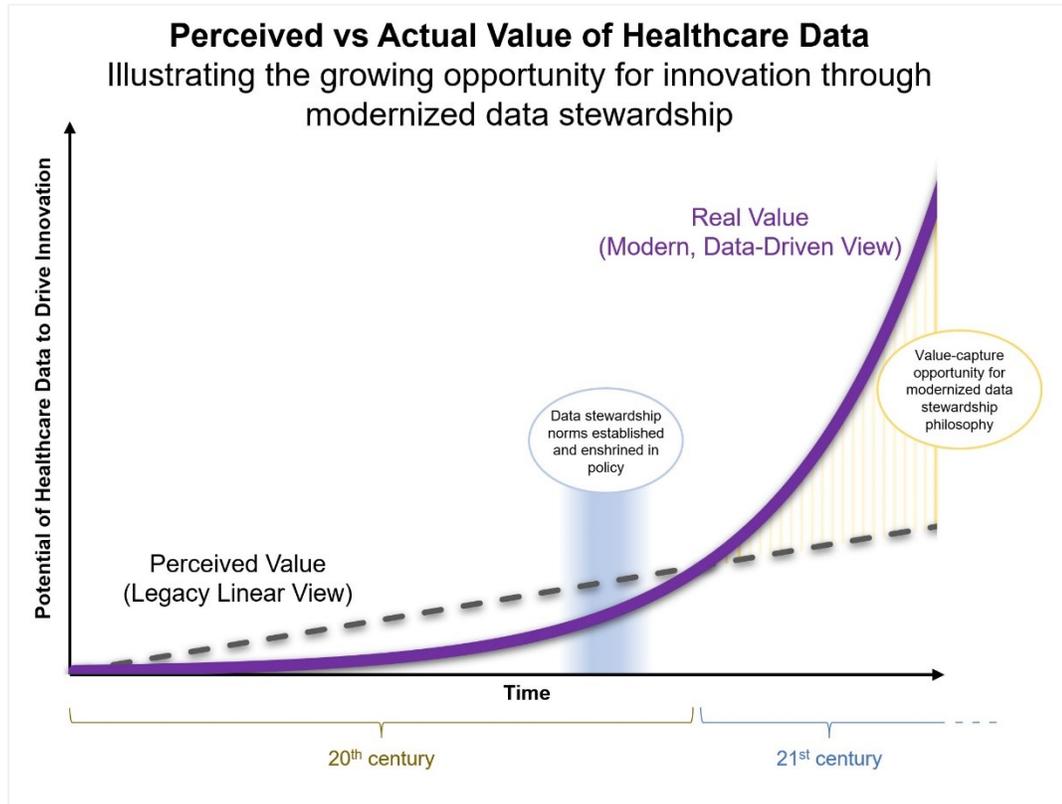

*Figure 1 - Contrasting Linear vs. Exponential Perceptions of Healthcare Data Value. This figure illustrates the divergence between linear perceptions of healthcare data value and its actual exponential capacity to drive innovation. The model highlights the growing gap between data value and legacy perception-based policy, emphasizing the misalignment between regulatory frameworks and the transformative potential of data-driven innovation in modern medicine.*

The other imperative is the elephant in the room—one the medical field has been conditioned not to see. Medical science in the 20th century achieved unprecedented progress, but it also entrenched deep inequalities. The recently coined term "financial toxicity" [12] has become a dominant feature of contemporary healthcare in the US, and expands the negative impact of inadequate healthcare to family budgets and on to all aspects of life. If there is a new way to lower the cost of healthcare, we have an ethical obligation to consider it.

Today, we stand at the edge of a paradigm shift—one that invites us to redefine our priorities and values. As healthcare evolves into a data-driven system powered by AI, software's value proposition has shifted from technical implementation to the strategic use of data.[13] By treating healthcare data as a shared resource and building governance structures that reflect that vision, we can catalyze the next generation of medical innovation and inspire the broader medical community to fully harness the transformative potential of emerging data technologies. The consequences of both action and inaction with respect to data stewardship are significant: either we will witness the power of neural networks integrating trillions of data points to revolutionize healthcare on all fronts within our lifetime, or we won't.

## The Growing Value of Data

"...sciences that involve human beings rather than elementary particles have proven more resistant to elegant mathematics...if that's so, we should stop acting as if our goal is to author extremely elegant theories, and instead embrace complexity and make use of the best ally we have: the unreasonable effectiveness of data." [1]

AI technology allows us now to process massive amounts of data at minute details.[14] When training such models, simple models given large amounts of data can perform as well or better than more complex algorithms trained on smaller data sets.[1]

Traditionally vital for documenting patient histories and retrospective analyses, healthcare data now holds transformative potential across numerous facets of medicine and healthcare management, contingent on effective data stewardship. An open healthcare data architecture would support advancements across all aspects of medicine, including innovation via machine learning, regulation, documentation and communication, outcome analysis, cost-benefit analysis, candidate identification, population health, genetic research, and emergency response.

Unlike hardware solutions, which take decades to diffuse and support global health through trickle-down processes[15], software-driven innovations scale at minimal cost and transmit globally at the speed of light. Expanding access to healthcare data would shift the focus of innovation investment from hardware-centric companies to software-centric solutions that connect individual patients, communities, and global health systems. Once acknowledged as a resource, we should reflect that it is unique among other types of economic resources such that healthcare data can be used and reused[1], potentially providing legacy value across interests, parties and time. Scientific studies, the backbone of innovation and evidence, could and *should* leverage millions of datasets routinely to identify patterns, interactions, and trends that have been invisible in traditional research models. Data-driven healthcare could and *should* foster greater transparency, equity, and efficiency in healthcare delivery.

# The Principle of Privacy

Privacy is a foundational principle in medical data management, essential for maintaining trust in the physician-patient relationship and upholding the integrity of healthcare systems. The American Medical Association (AMA) Code of Medical Ethics emphasizes that protecting patient confidentiality safeguards autonomy, prevents harm, and encourages open communication necessary for effective care (AMA Code of Medical Ethics, Opinion 3.2.1). The AMA asserts that safeguarding patients' privacy protects them from potential harm, ensures their autonomy, and encourages open communication necessary for effective care. The commitment to patient privacy is deeply embedded in medical ethics, legal requirements, and professional standards, reinforcing its role as a cornerstone of responsible data management.

Privacy protections exist for many reasons, addressing risks to individual patients as well as the public at large. Healthcare requires confidentiality, autonomy, and consent; protection against employment, insurance, and social discrimination; protection against identity theft and fraud, bias, and exploitation—all reasonable justifications for putting privacy protections in place.

Medical data policies have long treated privacy as the default solution for data protection—a logical approach when HIPAA and other foundational regulations were established in the 1990s, during the early integration of computing in healthcare. However, whether privacy-centric regulations still provide effective protection in today's digital landscape remains uncertain. At the same time, alternative approaches may better address privacy and patient protection while avoiding the unintended consequence of stifling innovation under outdated data management policies.

Medicine, computing, and computational power have changed drastically in the last few decades, and the cost and benefit of *privacy-at-all-costs data policy* should be reanalyzed light of modern technology. The *cost* of current rigid privacy frameworks includes past, present, and future opportunities missed across many facets of medicine, particularly in addressing critical challenges like the rising economic burden of healthcare. The perceived benefits also warrant reevaluation, as medical data is being aggregated regardless of patient awareness or consent by hospitals, companies, and organizations seeking to leverage it. Navigating patient interests with respect to privacy and innovation requires space for open discussion. *If* privacy is to remain paramount in medical data management, we must develop a contemporary rationale to support this through open dialog —one robust enough to support "acceptable" data aggregation strategies and withstand pressure from capital rich corporations and other powerful entities that will inevitably seek access to this data. Preparing for the reality of medical data as a resource demands proactive debate now, ensuring that future policies are both resilient and aligned with technological and ethical progress.

# Reevaluating ethical frameworks in data-driven healthcare

"Our efforts to seek transformation, while protecting our legacy models from disruption, may be antithetical."[16]

Ethical principles are essential for protecting patients and ensuring healthcare advancements align with societal values. The core tenets of biomedical ethics—beneficence, non-maleficence, autonomy, and justice—remain as relevant as ever, yet their expression must evolve in the context of modern data-driven healthcare[17] (see Table 1). While traditional data privacy and protection measures were once essential safeguards for upholding these ethical principles, they now risk obstructing innovation and limiting equitable access to care in ways that ultimately contradict those very principles.

The traditional practice of tightly controlling data access—creating silos—may unintentionally stifle innovation that could serve the greater public good. Beneficence, once focused solely on the immediate well-being of individual patients, must now extend to fostering an ecosystem of innovation that benefits patients on a broader scale. As the landscape evolves, protecting patients' rights must also include ensuring access to modern, data-driven technologies that lower costs and improve care.

Non-maleficence obligates us to protect patients from data misuse, raising a critical question: Are patients better served by legacy frameworks that impose broad legal liability on all data handlers, or by a targeted approach that enhances transparency and holds those who misuse data with malicious intent accountable? The cost of inaction—failing to update outdated data policies—is increasingly evident. The principle of "do no harm" must now encompass the risks of maintaining a restrictive status quo, which hinders advancements in care and perpetuates disparities.

Autonomy rightly prioritizes patient privacy, but it must also recognize patients' desire for groundbreaking, accessible innovations, cost-effective care, and transparent health systems—equally integral to justice. Balancing these priorities requires acknowledging the evolving landscape and embracing data-driven approaches that enhance healthcare delivery and patient outcomes without compromising ethical integrity.

*Table 1 - Evolution and future ambitions of medical ethics in healthcare*

| Ethical principle | 20th century medicine (privacy-centric, siloed) | 21st century medicine as understood today (data-driven, patient-centered) | Ambitions for 21st century medicine (transformational, innovation-driven) |
|---|---|---|---|
| **Autonomy** (patient rights, informed decision-making) | Strict privacy, limited patient involvement in data use | Patients have greater control over their data, opt-in models, shared decision-making | Patients as full partners in data ecosystems, benefiting from AI-driven insights, automated risk assessments, and personalized preventative care |
| **Beneficence** (maximizing benefits for patients) | One-size-fits-all treatments based on limited datasets | Personalized medicine and AI-driven diagnostics emerging but not yet widespread | Fully AI-integrated precision medicine, real-world data guiding treatments, expanded access to innovative therapies, and continuous learning health systems |
| **Nonmaleficence** (do no harm, risk mitigation) | Privacy-first approach prevents perceived harms from data exposure | Balanced privacy protections recognizing that lack of data sharing hinders care improvements | Regulatory frameworks that safeguard ethical data use while maximizing healthcare advancements, real-time monitoring of risks, and AI-assisted error prevention |
| **Justice** (fairness, access, equity) | Unequal data access, research limited to privileged institutions | Growing push for open data access to promote equity, but disparities persist | Truly democratized healthcare—open data fueling global innovation, reducing costs, increasing accessibility, and bridging healthcare disparities worldwide |

# Limitations of current data policies

"Despite having security standards such as HIPAA (Health Insurance Portability and Accountability Act), data breaches still happen on a daily basis. All various types of data breaches have a similar harmful impact on healthcare data, especially on patients' privacy." [6]

In the United States, medical data is bifurcated into research and clinical data, with both categories frequently constrained by HIPAA's stringent security mandates. In research, the NIH mandates sharing funded research data, a policy that conflicts with legal requirements to secure the same data.[7,18] Moreover, despite protocols and extraordinary costs spent to ensure data security, data breaches are not uncommon[6], and a significant lack of trust persists among stakeholders, further limiting healthcare data sharing.[19] These contradictions reveal how outdated frameworks hinder the potential of modern, data-driven healthcare.

Efforts to aggregate and utilize healthcare data often rely on centralized data storage (repositories) or federated networks. Centralized repositories streamline access but are expensive to maintain, operationally complex, and vulnerable to cyberattacks.[4] Conversely, federated networks decentralize data, reducing certain security risks, but introduce challenges such as data heterogeneity and the need for extensive coordination among stakeholders, which can be complex and resource-intensive.[20] Both approaches require substantial resources, and their scalability tends to increase linearly with the volume of data—an unsustainable model in an era of exponential data growth, underscoring the need for more scalable models. Furthermore, the lack of interoperability and standardization—evident in mismatched data formats and inconsistent coding—further compounds these limitations, making truly scalable solutions even more elusive.

These fragmented policies also limit patient involvement with their data. Current systems exclude individuals from knowledge or decisions about how their data is used, missing an opportunity to build trust through tiered consent or opt-in models.[21]

The economic costs of this stagnation are substantial. While industries like finance and retail have harnessed data to improve efficiencies and reduce costs, healthcare remains burdened by policies that constrain progress. Real-world evidence (RWE), a critical resource for scientific feedback, regulatory approvals, and post-market surveillance, remains underutilized due to these restrictions.[22]

Current policies also perpetuate inequities by favoring well-resourced institutions that can navigate the complex regulatory landscape. In academia, conducting interventional or observational research with human subjects requires teams of lawyers, coordinators, and support specialists—resources smaller institutions may lack. In industry, small startups—traditionally engines of innovation—struggle to compete with large companies whose entrenched business models resist transformation.

Lastly, medicine suffers from a lack of imagination. Few initiatives have greater potential to shape 21st century healthcare than data aggregation, yet proposals to rethink existing policies are often met with skepticism and resistance. Even ethical debates, which seek to examine moral frameworks, frequently ignore the evolving role of patient data, and fail to consider it is the vast majority of the global population who stands to benefit from its transformative potential.

## Opportunities for an Open Data Paradigm

The regulation of valuable resources has always adapted as their significance becomes clear, and healthcare data should be no exception. It is not impossible to imagine a different paradigm with medical data proactively treated as a resource to be managed for the public good. And the ideas here only represent the start of the critical conversation.

Instead of enforcing 20th century privacy models that create silos and restrict progress, policies could support a paradigm shift. They should be restructured to enhance accessibility, promote ethical use, and ensure accountability for misuse. The goal should be to build an innovation ecosystem that unlocks the potential of healthcare data while protecting patient trust—not by limiting access to all but by designing a system that balances transparency, responsibility, and innovation.

To achieve this, we should move toward a genuinely open healthcare data system where qualified data flows freely rather than being locked behind restrictive regulations. For example, anonymized healthcare records could automatically transition into the public domain after a qualification period triggered by factors such as a patient's death or a time-based embargo (e.g., 5 or 10 years after collection). This would ensure that patient data remains accessible for research, innovation, and policy planning without compromising the immediate privacy concerns of living individuals.

A real-time access model could also be established, allowing researchers, developers, and clinicians to retrieve anonymized data on demand, perhaps for a retrieval fee that covers costs. Rather than restricting access outright, this system should treat healthcare data as a public resource that remains open but is responsibly managed to prevent abuse. The revenue from retrieval fees could be reinvested into maintaining secure data infrastructures, funding medical research, or subsidizing healthcare costs for contributors.

Shifting the focus from privacy at all costs to responsible access and use would profoundly benefit medicine. First, such a system would eliminate silos rather than reinforce them, paving the way for a truly transformative era of innovation. Current frameworks default to restricting access, making it difficult for researchers and innovators to fully leverage healthcare data's power. A structured, open-access approach would create clear pathways for data transmission while maintaining necessary safeguards. Advances in technology already allow secure access models, and policies should reflect this reality rather than relying on outdated notions of secrecy.

Second, misuse of data—not responsible stewardship—should be penalized. The assumption that data can be perfectly anonymized no longer holds in an era of powerful computing and sophisticated reidentification techniques. With enough auxiliary data, even well-anonymized datasets can often be linked back to individuals. Rather than placing excessive restrictions on access in an unrealistic attempt to ensure perfect privacy while holding everyone who interacts with data as responsible for its use, policies should instead deter and punish actual misuse—whether that be unauthorized reidentification, discrimination, or unethical commercialization.

Third, this system should offer individuals choices regarding their data. While some may prefer to keep their data private indefinitely and be willing to pay the associated costs, others may prioritize accessing healthcare at reduced costs, opting out of additional fees tied to privacy protections. Beyond the direct economics of privacy, those who contribute to this societal resource could receive preferential access to data-driven healthcare solutions. Just as organ donation allows individuals to contribute to the greater good after death, a structured data donation system could enable voluntary participation in healthcare data-sharing initiatives, providing contributors with additional benefits.

Fourth, anonymization efforts should remain in place but be understood as imperfect. While de-identification is a valuable safeguard, it should not be relied upon as the sole means of protection. Instead, future data governance must prioritize transparency, accountability, and public oversight. Clear regulations, open reporting, and ethical review processes will ensure data is used for legitimate scientific and medical purposes. As Justice Brandeis said, "sunlight is the best disinfectant" [23]—transparency empowers trust, deters misuse, and builds a foundation for responsible data stewardship. This ideal stands in sharp contrast to the opaque and fragmented landscape that defines data governance today.

## Defining the New Ethics of Medical Data Stewardship

The new data stewardship redefines our collective responsibility toward health data in the 21st century. It prioritizes transparency over secrecy, shared value over proprietary control, and adaptive governance over rigid dogma. It calls for treating health data not merely as a byproduct of care, but as a shared public good—one to be protected, cultivated, and leveraged to advance science, equity, and patient well-being. This model rejects both unchecked exploitation and paralyzing fear. Instead, it seeks balance through accountable openness: open frameworks for innovation, accountable systems for protection, and inclusive dialogue to guide decision-making in a data-rich future.

## Conclusions

Healthcare is shaped not only by breakthroughs in science but by the conversations we allow ourselves to have. Fields like medical physics, which sit at the intersection of data and care, have a particular responsibility to speak plainly in moments of transformation. Sometimes leadership begins with stating what should be obvious: that it's not just acceptable—but essential—to talk openly about how norms must evolve. Progress depends not on perfect consensus, but on thoughtful, courageous discourse.

As healthcare transitions into an era of data driven innovation, data governance at government levels will determine whether innovation flourishes to benefit society or stalls under restrictive policies. We may borrow some perspective from history: our challenge of defining values in an uncharted paradigm shift is not unprecedented. In the late 19th century, Nikola Tesla and Thomas Edison battled over the future of electricity. Edison, with entrenched interests and financial backing, sought to maintain control over a proprietary, inefficient system—direct current (DC). Tesla, advocating for alternating current (AC), championed a model that was more scalable, efficient, and universally accessible. Despite resistance from powerful incumbents, the adoption of AC transformed global infrastructure and enabled modern technological expansion.

We now stand at a similar juncture in medicine. The foundation of evidence-based medicine—medical data—can either be exponentially accessible or locked in silos and used in shadows. The choices we make in structuring regulations, free markets, and government oversight will shape global health well into the 21st century, determining which solutions emerge and when—decisions that will have consequences for everyone reading this text.

Despite the urgency of this decision, our progress has been stalled—not by technological barriers, but by deeply ingrained social dogmas that have treated privacy as an absolute, precluding meaningful discussion. Privacy has been framed as unquestionable rather than as one of many competing values in healthcare, preventing proactive policymaking and leaving us unprepared for the realities of a data-driven medical future.

This moment calls for open, transparent conversations about our values. Balancing innovation, equity, and privacy requires engaging diverse stakeholders—including patients, researchers, policymakers, and industry leaders. By aligning regulation with modern capabilities and societal goals, we can unlock the full potential of healthcare data, empowering future generations with transformative tools for health and discovery.

**Funding statement:**

No funding was received for this work.

**Conflicts of interest:**

A.L.K. has served as a consultant for Boston Scientific. P.K. has financial relationships with Novartis, Bayer, Lantheus, Blue Earth, ConcertAI, GE Healthcare, Siemens/Cardinal Health, Pfizer, Merck, Johnson & Johnson, Clarity Pharmaceuticals, Curium, Astellas, and Telix Pharmaceuticals.